# Ti$^{4+}$ Substituted Magnesium Hydride as Promising Material for Hydrogen Storage and Photovoltaic Applications

R. Varunaa[1,2], S. Kiruthika[1,2] and P. Ravindran[1,2,3, a)]

[1]*Department of Physics, Central University of Tamil Nadu, Thiruvarur-610101.*
[2]*Simulation Center for Atomic and Nanoscale MATerials, Central University of Tamil Nadu, Thiruvarur-610101.*
[3]*Center for Material Science and Nanotechnology and Department of Chemistry, University of Oslo, Box 1033 Blindern, N-0315 Oslo, Norway*

[a)]Corresponding author: raviphy@cutn.ac.in

**Abstract.** In order to overcome the disadvantages of MgH$_2$ towards its applications in on-board hydrogen storage, first principle calculations have been performed for Ti (2+, 3+, and 4+) substituted MgH$_2$. Our calculated enthalpy of formation and H site energy implies that Ti substitution in Mg site reduces the stability of MgH$_2$ which improve the hydrogen storage properties and Ti prefers to be in +4 oxidation state in MgH$_2$. The bonding analyses through partial density of states, electron localization function and Bader charge of these systems confirm the existence of iono-covalent bonding. Electronic structure obtained from hybrid functional calculations show that intermediate bands (IB) are formed in Ti$^{4+}$ substituted MgH$_2$ which could improve the solar cell efficiencies due to multiple photon absorption from valence band to conduction band via IBs and converts low energy photons in the solar spectrum also into electricity. Further, our calculated carrier effective masses and optical absorption spectra show that Ti$^{4+}$ substituted MgH$_2$ is suitable for higher efficiency photovoltaic applications. Our results suggest that Ti$^{4+}$ substituted MgH$_2$ can be considered as a promising material for hydrogen storage as well as photovoltaic applications.

## INTRODUCTION

The researchers are developing new effective energy materials to address the energy and other environmental issues. Magnesium hydride (MgH$_2$) has been considered for hydrogen storage applications owing to its high hydrogen storage capacity, light weight, low cost, and abundance.[1] Its main limitations for on-board applications are the high decomposition temperature and the slow hydrogen sorption kinetics. Many studies report that titanium act as an efficient catalysts for hydrogen sorption, generally due to the high affinity toward hydrogen. Chengshang *et al*.[2] reported that the titanium additive added MgH$_2$ improves the hydrogen desorption temperature. Herein, we report the structural stability, electronic structure, chemical bonding, optical absorption spectra, and effective masses of MgH$_2$ and Ti substituted MgH$_2$ (MgH$_2$:Ti) to use for energy applications. Interestingly, we found that intermediate bands (IB) are formed in Ti$^{4+}$ substituted MgH$_2$ (MgH$_2$:Ti$^{4+}$). Several oxides and sulfides based IB solar cell materials have been identified recently[3] and here we report that IB can be formed in metal hydrides also by proper substitution.

## COMPUTATIONAL DETAILS

Density functional theory (DFT) calculations were performed using projector augmented wave potential with the exchange and correlation in the Perdew-Burke-Ernzerhof (PBE) formalism of DFT as implemented in Vienna *ab-initio* simulation package.[4] We have also employed Heyd-Scuseria-Ernzerhof (HSE06) hybrid functional with 25% Hartree-Fock exchange energy to obtain accurate band structure. The structural optimizations have been done until the force acting on all atomic sites were less than $10^{-3}$ eV/Å. An energy cutoff of 300 eV were used. The total energies were calculated by integration over a Monkhorst-Pack mesh of **k**-points in the Brillouin zone.

## RESULTS AND DISCUSSION

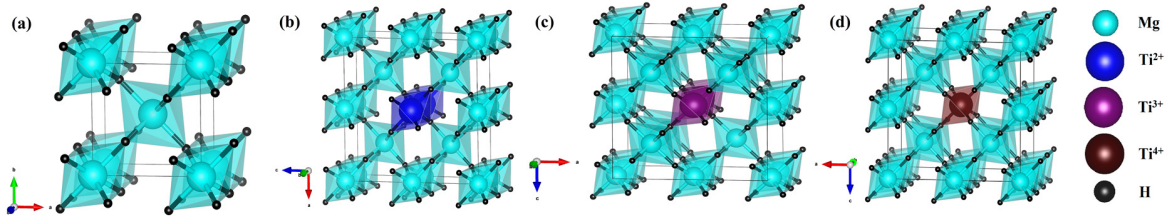

**FIGURE 1.** Crystal structure of (a) $MgH_2$, (b) $Mg_7TiH_{16}$ ($Ti^{2+}$), (c) $Mg_{13}Ti_2H_{32}$ ($Ti^{3+}$), and (d) $Mg_{14}TiH_{32}$ ($Ti^{4+}$).

In order to determine the stable oxidation state of Ti in Ti substituted $MgH_2$, we have modeled $MgH_2$:Ti system where Ti may have three possible oxidation states such as +2, +3, and +4. The calculated equilibrium lattice parameters ($a=b=4.4928$Å, $c=3.0037$Å) and volume (60.63Å$^3$) for $MgH_2$ are in good agreement with the experimental data. [5] At room temperature, $MgH_2$ crystallizes in the tetragonal (P4$_2$mnm) structure. The primitive unit cell of $MgH_2$ in Fig. 1(a) shows that each Mg atom is octahedrally coordinated to six H atoms. To construct $Ti^{2+}$, $Ti^{3+}$, and $Ti^{4+}$ substituted $MgH_2$, we have taken 2x2x1 (8 Mg and 16 H atoms), 2x2x2 (16 Mg atoms and 32 H atoms), and 2x2x2 (16 Mg atoms and 32 H atoms) super cell of $MgH_2$, respectively. For $Ti^{2+}$ substituted $MgH_2$, we have replaced one Mg with one Ti atom and the crystal structure of $Mg_7TiH_{16}$ is plotted in Fig. 1(b). To model $Ti^{3+}$ substituted $MgH_2$, a supercell is created with two Mg atoms are replaced by two Ti atoms as well as one Mg vacancy i.e. $Mg_{13}Ti_2H_{32}$ (see Fig. 1(c)). For $Ti^{4+}$ substituted $MgH_2$, we have replaced one Mg with one Ti atom and also one Mg vacancy is created i.e. $Mg_{14}TiH_{32}$ (see Fig. 1(d)). In Ti substituted $MgH_2$, each Mg is octahedrally coordinated with six H atoms and each Ti is also octahedrally coordinated with six H atoms.

**TABLE 1.** Calculated enthalpy of formation, H site energy, and band gap for pure $MgH_2$ and Ti substituted $MgH_2$.

| Compound | Enthalpy of formation, $\Delta H_f$ (kJ/mol.H$_2$) | H site energy (eV) | | Band gap (eV) GGA | HSE |
|---|---|---|---|---|---|
| $Mg_2H_4$ | -52.77 (-45.19)[1] (-60.3) [6] | 1.203 | | 3.69 (3.74)[7] | 4.82 |
| $Mg_7TiH_{16}$ ($Ti^{2+}$) | -45.67 (-44.8) [8] | 0.693 | | metal | - |
| $Mg_{13}Ti_2H_{32}$ ($Ti^{3+}$) | -42.63 | 0.558 | | 0.29 | - |
| $Mg_{14}TiH_{32}$ ($Ti^{4+}$) | -29.79 | 0.720 | $E_g1$ | 0.64 | 1.59 |
| | | | $E_g2$ | 1.91 | 1.44 |
| | | | $E_g3$ | 0.18 | 0.15 |

From the calculated enthalpy of formation and H site energy (see Table 1), we have found that Ti substituted $MgH_2$ systems are less stable than pure $MgH_2$. Our calculated enthalpy of formation of $MgH_2$ and Ti substituted $MgH_2$ are in good agreement with other theoretical/experimental observations. Hence the Ti substitution reduces the stability of $MgH_2$, as a result, the dehydrogenation temperature will be lower for Ti substituted $MgH_2$ than pure $MgH_2$. We have found that the H site energy is smaller for Ti substituted $MgH_2$ compared with pure $MgH_2$. This result also implies that H can be desorbed easily in $MgH_2$:Ti than from pure $MgH_2$. Moreover, among the various configurations considered in the present study, the H site energy is more for $Ti^{4+}$ substituted $MgH_2$ than $Ti^{2+}$ or $Ti^{3+}$ substituted $MgH_2$ indicating that Ti in the +4 oxidation state is energetically more favorable.

The calculated total density of states (DOS) for all these systems are displayed in Fig. 2(a) and from this we found that $MgH_2$ is an indirect band gap insulators. Our results show that Ti substitution in $MgH_2$ reduces the band gap. The appearance of a small energy gap confirms the semiconducting nature of the Ti substituted $MgH_2$. The band structure of $MgH_2$ and $MgH_2$:$Ti^{4+}$ calculated using PBE and HSE06 functional are shown in Fig. 2(b,c) and 3(c,d), respectively and the corresponding band gaps are listed in Table 1. Due to the high value of bandgap in $MgH_2$ it will be useful to harvest solar energy in UV range only. However the formation of intermediate band states by Ti substitution opens up the possibility of harvesting solar energy in the visible and other lower energy region in the solar spectrum and hence one can expect that the energy conversion efficiency can be improved if one use such intermediate band gap material.

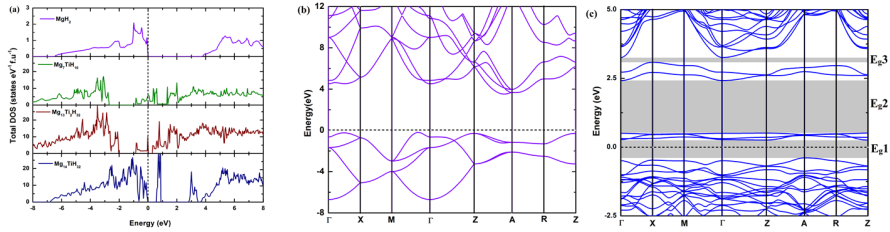

**FIGURE 2.** Calculated (a) total DOS for MgH$_2$:Ti and band structure of (b) MgH$_2$ and (c) MgH$_2$:Ti$^{4+}$ using PBE functional.

The partial DOS of MgH$_2$ (see Fig. 3(a)) shows that the contribution of H is higher than that of Mg in the valence band (VB) and each Mg atom donates around 1.59 electrons to the neighbouring H atoms (see Table 2) which indicates the presence of ionic bonding. Moreover, the H-*s* states are hybridized with Mg-*p* state in the whole VB which hints the existence of covalent bonding as well. For MgH$_2$:Ti$^{4+}$, the partial DOS are plotted in Fig. 3(b). In this system, it is interesting to note that the Ti-*d* states form an intermediate band states in the middle of the band gap. The electronic states from Ti, Mg, and H are energetically degenerate in the energy range -2 eV to VB maximum indicating the presence of the noticeable covalent bond between them. The majority of the Ti-*d* states in the VB are well separated and present between -4 eV and -2 eV. Most of the Mg-*s* electrons are transferred to H site in MgH$_2$:Ti$^{4+}$ also which can be seen from the calculated Bader charge (see Table. 2) indicating substantial ionic bonding. Hence the bonding in MgH$_2$:Ti$^{4+}$ can be considered as mixed iono-covalent character.

**TABLE 2.** Calculated Bader effective charges (BC) for the constituents.

| Compound | MgH$_2$ | | Mg$_{14}$TiH$_{32}$ | | |
|---|---|---|---|---|---|
| atom | Mg | H | Mg | Ti | H |
| BC (*e*) | 1.5869 | -0.7934 | 1.5847 | 1.8705 | -0.7518 |

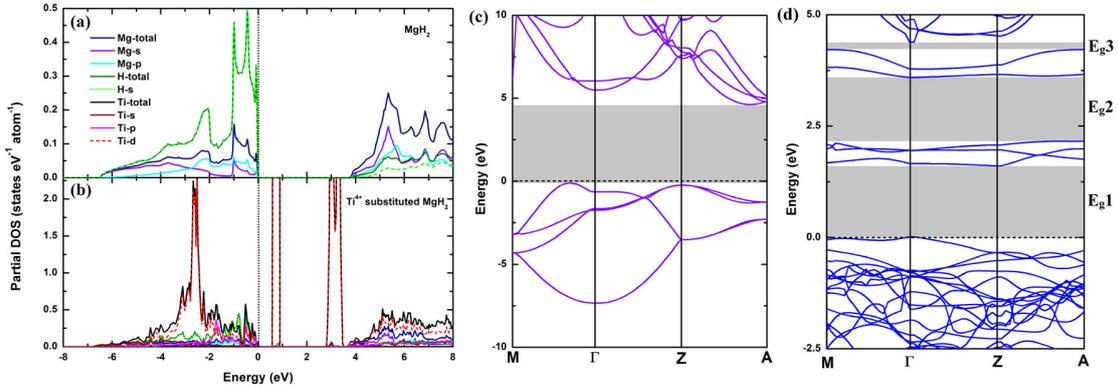

**FIGURE 3.** (a,b) Partial DOS (PBE) and (c,d) band structure (HSE06) of MgH$_2$ and Ti$^{4+}$ substituted MgH$_2$.

The electron localization function (ELF) plot for MgH$_2$ (see Fig. 4(a)) indicating that the charges are accumulated at H site and depleted at Mg site. This indicate the charge transfer from Mg site to H site suggesting the presence of ionic bonding between them. Moreover, the noticeable charges present in the interstitial region indicating the presence of noticeable covalency as well. The ELF plot for Ti$^{4+}$ substituted MgH$_2$ is displayed in Fig. 4(b). One can notice that the strong covalent bonding of Ti atom with the surrounding H atoms is expressed by the significant amount of shared charges in the plane. Moreover, the charge transfer from Mg to neighboring H atoms indicates the presence of ionic bonding between Mg and H.

It is interesting to note that Ti substitution forms an intermediate bands in the band gap. So, it can be used for photovoltaic applications since the multi-band gap materials offer the possibility to improve the solar cells efficiency than the traditional single band gap solar cell materials. In order to evaluate carrier mobility, we have calculated carrier effective masses as it is inversely proportional to the carrier mobility. For $MgH_2$, the calculated effective mass of hole along M-Γ is 0.34 $m_0$ whereas electron effective mass along Z-A direction is 0.67 $m_0$. For $Ti^{4+}$ substituted $MgH_2$, the hole effective mass along Z-A direction is 1.01 $m_0$ and electron effective mass along Γ-Z direction is 0.59 $m_0$. Present results indicate that Ti substitution reduces electron effective mass and hence will improve the electron transport. The calculated optical absorption spectrum of $MgH_2$ and $Ti^{4+}$ substituted $MgH_2$ using PBE functional followed by a rigid energy shift using the improved bandgap value from HSE06 to take into account of band gap underestimation by PBE functional is shown in Fig. 4(c). For $MgH_2$, the optical absorption starts at ~4.8 eV and increases with photon energy. In the case of $Ti^{4+}$ substituted system, the absorption starts at ~1.6 eV and increases with photon energy over the range of visible light. This indicates that $Ti^{4+}$ substituted $MgH_2$ may be a potential material for IB solar cells.

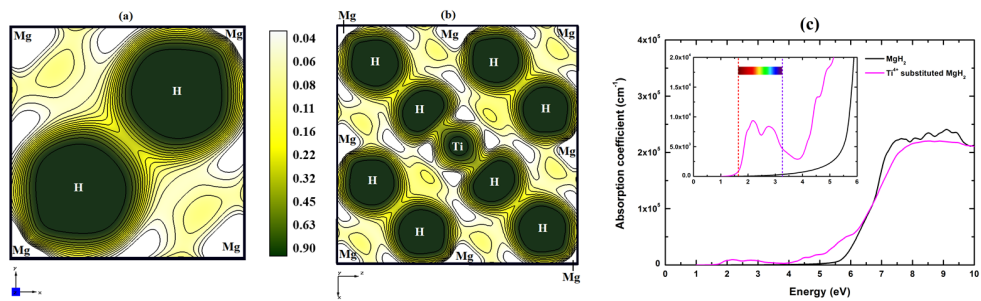

**FIGURE 4.** Calculated ELF plot for (a) $MgH_2$, (b) $Ti^{4+}$ substituted $MgH_2$ and (c) their absorption spectra.

## CONCLUSIONS

In conclusions, our total energy studies show that the stability of $MgH_2$ is reduced by Ti substitution in $MgH_2$ since our calculated $\Delta H_f$ and H site energy of Ti substituted $MgH_2$ systems are lower than those for $MgH_2$. Further, present results reveal that Ti prefers to be in +4 oxidation state in $MgH_2$:Ti. Electronic structure calculations show that $MgH_2$ has insulating behavior whereas $Ti^{4+}$ substituted $MgH_2$ has semiconducting behavior with an intermediate bands in the band gap region. The chemical bonding analyses through partial DOS, ELF and Bader charge confirm the presence of iono-covalent nature in these systems. Effective mass and optical absorption calculations show that $Ti^{4+}$ substituted $MgH_2$ could be a potential material for IB solar cells. The present study suggests that $Ti^{4+}$ substituted $MgH_2$ will be a promising material for on-board hydrogen storage applications and also can be used as an active material in the intermediate band gap solar cells.

## ACKNOWLEDGMENTS

The authors are grateful to the Department of Science and Technology, India for the funding support via Grant No. SR/NM/NS-1123/2013 and the Research Council of Norway for computing time on the Norwegian supercomputer facilities (Project No: NN2875K).